\documentclass[12pt]{emulateapj}

\def\ngc#1{\hbox{NGC\,#1}}
\def\U{$U$}
\def\B{$B$}
\def\V{$V$}
\def\I{$I$}
\def\umv{\hbox{\it U--V\/}}
\def\bmi{\hbox{\it B--I\/}}
\def\bmv{\hbox{\it B--V\/}}
\def\vmi{\hbox{\it V--I\/}}

\def\deg{${}^\circ$}

\def\ltsim{ \,{}^<_\sim\, }

\shorttitle{On the stellar content of Carina} 
\shortauthors{Bono et al.}

\begin{document}
\title{On the stellar content of the Carina dwarf spheroidal galaxy\altaffilmark{1}}  

\author{
G.~Bono\altaffilmark{2,3},
P.~B.\ Stetson\altaffilmark{4,18},
A.~R.\ Walker\altaffilmark{5},
M.~Monelli\altaffilmark{6},
M.~Fabrizio\altaffilmark{2},
A.~Pietrinferni\altaffilmark{7},
E.~Brocato\altaffilmark{7},
R.~Buonanno\altaffilmark{2,8},
F.~Caputo\altaffilmark{3},
S.~Cassisi\altaffilmark{7},
M.~Castellani\altaffilmark{3},
M.~Cignoni\altaffilmark{9,10},
C.~E.\ Corsi\altaffilmark{3},
M.~Dall'Ora\altaffilmark{11},
S.~Degl'Innocenti\altaffilmark{12,13},
P.~Fran\c{c}ois\altaffilmark{14},
I.~Ferraro\altaffilmark{3},
G.~Iannicola\altaffilmark{3},
M.~Nonino\altaffilmark{15}, 
P.~G.\ Prada Moroni\altaffilmark{12,13},
L.~Pulone\altaffilmark{3}, 
H.~A.\ Smith\altaffilmark{16},
F.~Thevenin\altaffilmark{17}
}

\altaffiltext{1}{Based on images collected with the MOSAICII camera available 
at the CTIO 4m Blanco telescope, La Serena; (2003B-0051, 2004B-0227, 2005B-0092, 
P.I.: A.R. Walker) and in part with the WFI available at the 2.2m MPG/ESO telescope 
(A064.L-0327) and with images obtained from the ESO/ST-ECF Science Archive Facility.}

\altaffiltext{2}{Dipartimento di Fisica, Universit\`a di Roma Tor Vergata, 
via della Ricerca Scientifica 1, 00133 Rome, Italy; bono@roma2.infn.it}
\altaffiltext{3}{INAF--OAR, via Frascati 33, Monte Porzio Catone, Rome, Italy}
\altaffiltext{4}{DAO--HIA, NRC, 5071 West Saanich Road, Victoria, BC V9E 2E7, Canada} 
\altaffiltext{5}{NOAO--CTIO, Casilla 603, La Serena, Chile}
\altaffiltext{6}{IAC, Calle Via Lactea, E38200 La Laguna, Tenerife, Spain}
\altaffiltext{7}{INAF--OACTe, via M. Maggini, 64100 Teramo, Italy}
\altaffiltext{8}{ASI--Science Data Center, ASDC c/o ESRIN, via G. Galilei, 00044 Frascati, Italy}
\altaffiltext{9}{Dipartimento di Astronomia, Universit\`a di Bologna, via Ranzani 1, 40127 Bologna, Italy}
\altaffiltext{10}{INAF--OAB, via Ranzani 1, 40127 Bologna, Italy}
\altaffiltext{11}{INAF--OACN, via Moiariello 16, 80131 Napoli, Italy}
\altaffiltext{12}{Dept. of Physics, Univ. Pisa, Largo B. Pontecorvo 2, 56127 Pisa, Italy}
\altaffiltext{13}{INFN, Sez. Pisa, via E. Fermi 2, 56127 Pisa, Italy}
\altaffiltext{14}{Observatoire de Paris-Meudon, GEPI, 61 avenue de l'Observatoire, 75014 Paris, France}
\altaffiltext{15}{INAF--OAT, via G.B. Tiepolo 11, 40131 Trieste, Italy}
\altaffiltext{16}{Dept. of Physics and Astron. Michigan State University, East Landsing, MI 48824, USA}
\altaffiltext{17}{Observatoire de la Cote d'Azur, BP 4229, 06304 Nice, France}
\altaffiltext{18}{Visiting Astronomer, Cerro Tololo Inter-American Observatory, 
National Optical Astronomy Observatories, operated by AURA, Inc., under cooperative 
agreement with the NSF.}

\date{\centering drafted \today\ / Received / Accepted }

\begin{abstract}
We present deep, accurate and homogeneous multiband optical (\U,\B,\V,\I)
photometry of the Carina dwarf spheroidal galaxy, based on more than 4,000
individual CCD images from three different ground-based telescopes.  Special
attention was given to the photometric calibration, and the precision for the
\B, \V, and \I bands is generally better than 0.01 mag.  We have performed
detailed comparisons in the \V, \bmv\ and \V, \bmi\ color-magnitude diagrams
(CMDs) between Carina and three old, metal-poor Galactic Globular Clusters
(GGCs, M53, M55, M79).  We find that only the more metal-poor GCs (M55,
[Fe/H]=--1.85; M53, [Fe/H]=--2.02 dex) provide a good match with the Carina
giant branch.  We have performed a similar comparison in the \V, \vmi\ CMD with
three intermediate-age clusters (IACs) of the Small Magellanic Cloud
(Kron~3, NGC~339, Lindsay~38). We find that the color extent of the SGB of the
two more metal-rich IACs (Kron~3, [Fe/H]=--1.08; NGC339, [Fe/H]=--1.36 dex) is
smaller than the range among Carina's intermediate-age stars. Moreover, the
slope of the RGB of these two IACs is shallower than the slope of the
Carina RGB. However, the ridge line of the more metal-poor IAC (Lindsay~38,
[Fe/H]=--1.59 dex) agrees quite well with the Carina intermediate-age stars. 
These findings indicate that Carina's old stellar population is metal-poor and
appears to have a limited spread in metallicity ($\Delta$[Fe/H]=0.2--0.3 dex). 
The Carina's intermediate-age stellar population can hardly be more metal-rich
than Lindsay~38 and its spread in metallicity also appears modest.  We also find
that the synthetic CMD constructed assuming a metallicity spread of 0.5~dex for
the intermediate-age stellar component predicts evolutionary features not
supported by observations.  In particular, red clump stars should attain colors
that are redder than red giant stars, but this is not seen.  The above results
are at odds with recent spectroscopic investigations suggesting that Carina
stars cover a broad range in metallicity ($\Delta$[Fe/H]$\sim$1--2 dex).  We
also present a new method to estimate the metallicity of complex stellar systems
using the difference in color between the red clump and the middle of the RR
Lyrae instability strip.  The observed colors of Carina's evolved stars indicate
a metallicity of [Fe/H]=--1.70$\pm$0.19 dex, which agrees quite well with
spectroscopic measurements.
\end{abstract}

\keywords{galaxies: dwarf --- galaxies: stellar content --- 
stars: evolution --- stars: fundamental parameters}

\maketitle

\section{Introduction}

Dwarf galaxies play a fundamental role in several astrophysical 
problems. Current cosmological simulations predict dwarf satellite 
populations significantly larger than the number of dwarfs observed near
giant spirals like the Milky Way and M31. This discrepancy is called the 
``missing satellite problem'' and is a challenge to the currently most popular 
cosmological model: the $\Lambda$ Cold Dark Matter paradigm 
\citep{kly99,moore06,mandau08}. However, it has not yet been 
established whether this discrepancy is due to limitations in the theoretical 
modeling or to observational bias at the faint end of the dwarf galaxy luminosity 
function \citep{kli09,kop09,krav10}.

Moreover, detailed analyses of the distribution of dwarf galaxies in multi-dimensional 
parameter space indicate that they follow very tight relations. In particular, \citet{prada02}
suggested that Local Group (LG) dwarf galaxies show strong 
correlations (Fundamental Line - FL) among mass-to-light (M/L) ratio, surface brightness,
and metallicity. They explained the correlation between M/L ratio and metallicity 
using a simple chemical enrichment model in which the hot metal-enriched gas, 
at the end of the star-formation epoch, is lost via continuous galactic winds 
\citep{maclow99}. In a recent investigation \citet{woo09} reached
similar conclusion, i.e., LG dwarf galaxies follow a one-parameter relation 
driven by the total stellar mass. They also found that dwarf spheroidals (dSphs) 
appear to be, at fixed stellar mass, systematically more metal-rich than dwarf 
irregulars (dIs, see also \citet{mateo98}). Futhermore, the FL of dIs in a 
five-dimensional parameter (total space mass, surface brightness, rotation velocity, 
metallicity, star formation rate) is linear and tight, and extends the 
scaling relations of giant late-type galaxies into the low-mass regime. 
On the other hand, the FL of dSphs as defined in a four-dimensional parameter 
space (total mass, surface brightness, rotation velocity, 
metallicity) is also linear and very tight, but it does {\it not\/} lie on an 
extrapolation of  the scaling relations of giant early-type 
galaxies.
Iron and heavy-element abundances, the spread in these quantities, and the star-formation 
history (SFH) of LG dwarf galaxies, are crucial observational constraints on the physical
mechanisms responsible for the scaling relations described above, and on their 
implications for different cosmological models \citep{orb08}.   
Spectroscopic investigations based on high-resolution spectra of bright red giants (RGs) 
in several LG dSphs indicate a large spread in iron abundance \citep{shet01,shet03}.
This is also supported by metallicity estimates from the CaII triplet method
\citep{bat08}. In this context, the Carina dSph is particularly relevant:
({\em i}) It is relatively close and its central density is modest ($\rho$=0.17$M_\odot/pc^3$, 
\citet{mateo98}) so it is easily resolved with ground-based telescopes \citep{mou83}. 
({\em ii}) It is {\it the\/} LG dSph most clearly showing multiple, widely separated star 
formation episodes(\citet{mig90,mig97}; \citet{smec94,smec96};
\citet{hur98}; \citet{her00}; \citet{har01}; \citet{dol02}; \citet{riz03}) with a tail 
that might extend to less than one Gyr ago (\citet{mon03}). 
({\em iii}) It is also a benchmark to constrain the pulsation properties of old (RR Lyrae, 
\citet{saha86}) and intermediate-age (dwarf Cepheids, \citet{matal98}; Anomalous Cepheids, 
\citet{ora03}) variable stars.
({\em iv}) High-resolution spectra are available for a sample of ten bright RGs. 
The mean metallicity  is [Fe/H]=--1.69 with $\sigma$=0.51 dex \citep{koch08}.  
Independent measurements by \citet{shet03}, using high resolution spectra 
for five bright RGs, provided a mean metallicity of [Fe/H]=--1.64 and $\sigma$=0.2 dex.
({\em iv}) Medium-resolution calcium triplet measurements are also available for
a large sample (437) of RG stars \citep{koch06}; their metallicity distribution shows a peak 
at \hbox{[Fe/H]=--1.72$\pm$0.01} (metallicity scale by Carretta \& Gratton 1997) and the 
metallicity ranges from $\sim$--2.5 to $\sim$--0.5 dex. Using the same spectra, but 
different selection criteria (364 candidate Carina stars), Helmi et al.\ (2006) found the 
same peak in the metallicity distribution ([Fe/H]=--1.7$\pm$0.1 dex) and metallicities 
ranging from $\sim$--2.3 to $\sim$--1.3 dex (see their Fig.~2). 

Deep, accurate photometric investigations indicate an old (13 Gyr) stellar population and  
several distinct intermediate-age populations (2--6 Gyr) together with a blue plume of younger 
stars. The occurrence of both old and intermediate-age populations is also indicated by two 
distinct samples of core helium burning stars:  an old, wide HB and a separate red clump (RC).
The shape and thickness of the RGB ($\Delta$\bmv$\sim$0.02 for 
20.5$\lesssim$$V$$\lesssim$20.75 mag) suggest that the old and intermediate-age 
stellar populations have a minimal spread in chemical composition.      
This seems inconsistent with the spectroscopic observations that 
indicate a metallicity distribution possibly reaching extreme values near --3.1 and 
+0.1~dex \citep[scale]{zinn84}, or near --2.8 and --0.2~dex \citep[scale]{car97}. 
 
Despite the criticisms raised by \citet{koch06}, conclusions based on purely photometric 
indicators are hampered by only two major factors: 
({\em i}) A decrease in photometric precision when moving from bright to faint RG stars
might produce a spread in color mimicking a spread in metal 
content. In particular, Carina's stellar density is very low,
requiring that photometry be obtained with multi-chip CCD cameras or with many
different pointings of a single CCD.
A high degree of consistency in the photometric zero-points across the
field is essential to avoid spurious broadening of the RG and Main Sequence (MS) loci. 
({\em ii}) Due to its low surface density, large extent, and  
low Galactic latitude (--22\deg), the Carina sample is 
contaminated by foreground field stars. Their colors, unfortunately, are 
similar to the Carina RGB, once again mimicking a broadening in color due to a spread
in age/metallicity.

Our group is involved on a long-term investigation of the stellar
populations in the Carina dSph. In particular, we have assembled our own and archival
{\it UBVI\/} images 
covering the entire body of the galaxy. In this investigation we focus on
comparing Carina with old and intermediate-age template clusters.
We also present a new method for estimating the metal content of complex stellar
systems  from the difference in color between the middle of the RR Lyrae
instability strip and the peak of RC stars.

\section{Observations and data reduction}
The data were collected in various observing runs between 1992 December and 2005
January. They include images from three telescopes: the CTIO 1.5m
telescope with a single Tektronix2K CCD, the CTIO 4m Blanco telescope with the
MosaicII camera, and the ESO/MPG 2.2m telescope with the Wide Field Imager
camera (both proprietary and archival data). Details of the
observations will be presented in a future paper. The data presented here
represent 4,152 individual CCD images with essentially complete coverage of the
central regions of Carina ($\approx40\arcmin\times55\arcmin$, lacking only those
regions obliterated by bright foreground stars), and some sampling of the
galaxy's halo out to an extreme radius of $\sim$108\arcmin.  They were obtained
in four photometric bands ($B$ and $I$ with the 1.5m telescope, and {\it UBVI\/}
with both the 4m and 2.2m telescopes).

The data were reduced using the \hbox{DAOPHOT/}\hbox{ALLFRAME} package
\citep{stet87,stet94}. Individual PSFs were produced for each chip of every
exposure, using semi-automated routines to select bright, isolated
PSF stars. Subsets of the data from each observing run were first
reduced separately. Finally, everything was merged together and a single run of
ALLFRAME was adopted to reduce all the images of the center of
Carina, with separate reductions for non-overlapping outlying fields. Because
of the multiple chips and pointings required to cover the galaxy, no star
appears in every image; any given star may have up to 17 calibrated
measurements in $U$, 156 in $B$, 207 in $V$, and 70 in $I$. A total
of 205,338 individual stars were catalogued and measured. Among these, 72,595
have photometric measurements in all four filters, and 129,230 have at
least $V$ and either \bmv\ or \vmi. The remainder, either extremely faint or
located near the periphery of our coverage, have astrometry and 
instrumental magnitudes only.   

The Carina data were contained within 206 individual datasets, where a dataset
is essentially the totality of data obtained with one CCD from one night of
observing. These, along with 1,331 datasets from other nights and other
telescopes, were calibrated to the current version of Stetson's (2000)
photometric standard system, which is believed to be equivalent to that of
\citep{land92} to well under 0.01~mag in each of $B$, $V$, and $I$ (see also
\citet{stet05}). The $U$ photometric bandpass is more problematic, and in this
paper we will employ the $U$-band data only for qualitative and relative
comparisons.

\section{Results and discussions}

For a robust discrimination of candidate Carina members from
foreground field stars and background galaxies we adopted the \umv,\bmi\ 
color-color diagram (C-CD). The top panel of Fig.~1 shows that
this is effective for distinguishing candidate Carina and field stars. 
Carina RGs (\umv$\ge$0.6, \bmi$\ge$1.4) attain, at fixed \bmi, 
bluer \umv\ colors than field stars. The difference is mainly caused by a
difference in the mean metallicity between Carina and the more metal-rich
Galactic disk stars, but the difference in surface gravity between galaxy giants
and foreground dwarfs may also play a role.  In contrast, hot horizontal-branch (HB) stars and
relatively young MS stars in Carina (\bmi$\lesssim$0.4) are relatively free from
field star contamination, since they are bluer than the thick
disk and halo turnoffs; compact background galaxies represent the principal
source of contamination among these bluer objects.  Fortunately, this
C-CD is also a good diagnostic to separate candidate Carina stars from
blue background galaxies since the latter show, at fixed \bmi, systematically
bluer \umv\ colors than stars of any luminosity class or metallicity (apart
from white dwarfs, which are not an issue here).

Based on the above empirical evidence, we performed a series of
tests to identify candidate Carina stars using various color-magnitude
diagrams (CMDs) together with the C-CD.  The solid-outlined box in the
top panel of Fig.~1 shows the final selection, while the bottom panels show the
$V$,\bmv\ CMD of the total sample (left), the candidate Carina stars (middle), and
the candidate non-member objects (right). Note that after the selection there
remain some field stars with colors similar to 
Carina's RGB ( $V$$\le$21.5, \bmv$\sim$0.4-0.6; middle panel). However, RG stars 
are distributed along a very narrow sequence over the entire magnitude range 
(18$\le$$V$$\le$23), and the number of field stars lying near that RGB in both 
color and magnitude is appreciably reduced. The same statement applies to the
intermediate-age red clump (RC: $V$$\sim$20.5, \bmv$\sim$0.65) and to the old HB 
($V$$\sim$20.75, \bmv$\sim$0.65).
These evolutionary phases are characterized by very 
narrow distributions in either magnitude or color or both. The improvement 
compared to data already available in the literature is due to smaller photometric 
uncertainties ($\sigma_{\bmv}$$\le$0.02 for $V$$\le$21, see error bars in Fig.~1).        
Plain evolutionary arguments \citep{mon03,zocc03} 
suggest that small dispersions in magnitude and color in these evolutionary 
phases imply a negligible spread in chemical composition. This finding supports 
the results based on B,V,I photometry of Carina RG stars obtained by \citet{riz03}. 
The presence of a few stars in the right panel with magnitudes and 
colors typical of red HB ($V$$\sim$20.75, \bmv$\sim$0.4) and faint subgiant branch (SGB) stars
($V$$\sim$23, \bmv$\sim$0.5) demonstrates that our C-CD selection is 
imperfect---especially at intermediate colors---and some real Carina members
have probably been erroneously rejected. However, in these photometric selections 
it is more important to keep probable non-members out than to keep possible 
members in \citep{cal09}. 

To provide robust constraints on Carina's stellar content, we compared it with stellar
systems characterized by different ages and metal abundances. In particular, to 
constrain the old stellar population we selected three globular clusters (GCs)---M79 = \ngc{1904}, M55 = \ngc{6809}, 
and M53 = \ngc{5024}---with metal abundances ranging from \hbox{--1.64}$\pm$0.15 to --2.02$\pm$0.15,
and low foreground reddenings (E(\bmv)$\le$0.09, see Table~1). The optical {\it BVI\/}
photometry of these GCs is also on the \citet{stet05} system.
The accuracy of the photometry is certainly better than our knowledge of the 
foreground reddening toward these clusters and Carina.  The $V$, \bmv\ (left panels) 
and $V$, \bmi\ (middle panels) CMDs of the selected GCs are shown in Fig.~2, together
with their individual ridge lines (red curves). We also selected three
intermediate-age clusters (IACs) belonging to the Small Magellanic Cloud
(SMC)---Kron~3, \ngc{339}, Lindsay~38---with similar ages and iron 
contents ranging from --1.08$\pm$0.12 to --1.59$\pm$0.10 dex 
\citep[see Table~1]{glatt09}. 
The photometry for these clusters is from ACS@HST data transformed 
into the $V$, $I$ Johnson-Cousins bands using relations by 
\citet{sir05}. These clusters are {\it not\/} included
in the homogenous photometry project (the original sources of the data are 
listed in Table~1), so we are unable to independently confirm the precision 
of the photometry or the accuracy of the calibrations.  Data plotted in the 
right panels of Fig.~2 display the $V$,\vmi\ CMDs for these clusters together with 
their ridge lines and the outlines of their RCs.

To compare the GCs with the old stellar populations in Carina, we arbitrarially 
shifted their ridge lines in magnitude and color to 
provide a good fit with the SGB (23.25$\lesssim$$V$$\lesssim$23.75, 
0.45$\le$\bmv$\le$0.60) in Carina.     
Then we inferred the difference in distance and in reddening between 
Carina ($\mu$=20.15, E(\bmv)=0.03; \citet{ora03,piet09}) and the individual GCs. The distances and the cluster reddenings we found   
following this approach agree, within the errors, with similar estimates 
available in the literature (see columns 4 and 5 in Table~1).  
The top and the middle left panels of Fig.~3 show that the GCs M79 and M55 
provide a poor match to Carina, since hot HB stars are systematically brighter 
than Carina HB stars. On the other hand, both the GB and the HB of the more 
metal-poor GC (M53) match Carina quite well. 
The same outcomes apply if the comparison is performed using 
the $V$, \bmi\ CMDs (see the middle panels in Fig.~3). The extinction in the 
{\it BVI\/} bands was estimated using the empirical relations provided by 
\citet{mccall04}.  

The same approach was followed to compare the IACs with Carina stellar populations. 
The ridge lines were shifted in magnitude and color to provide a good fit to
the lower envelope of intermediate-age SGB 
(22.5$\lesssim$$V$$\lesssim$23.0, 0.45$\lesssim$\vmi$\lesssim$0.80) stars in Carina.  
The distances that we estimate are larger than those 
estimated by \citet{glatt09} (see Table~1) using 
isochrones. However, the current distances agree, within the errors, with 
SMC distances available in the literature \citep{roman09}. The same 
outcome applies for the cluster reddenings (see the right panels in Fig.~3). 
This comparison between Carina and the IACs revealed the following: 
({\em i}) the color extents of the SGBs of the the two more metal-rich clusters (Kron~3 
and \ngc{339}; i.e., the length of the horizontal sequence between the upper MS
and the lower GB) are smaller than the color range covered by Carina's intermediate-age 
SGB stars; ({\em ii}) the slope of the RGB in the two more metal-rich clusters 
(Kron~3, \ngc{339}) is shallower than the observed slope of Carina's RGB;     
({\em iii}) the ridge line of Lindsay~38 is a fairly good match to Carina's
intermediate-age population. 
({\em iv}) the RC contour of Kron~3 agrees with Carina's RC, while for 
the other two clusters the clumps are either slightly redder and brighter (\ngc{339}) 
or marginally fainter (Lindsay~38).  

These comparisons indicate that Carina's old population is relatively 
metal-poor, and its spread in metallicity is modest: at most 0.2--0.3 dex. 
Carina's intermediate-age population seems to have the same metallicity as
Lindsay~38, or perhaps slightly lower.  
The spread in metal abundance in this stellar component also seems
very limited, since the RGBs of the two more metal-rich IACs
are considerably redder than the red envelope of Carina's RGB.
The lack of accurate 
photometry for metal-poor ([Fe/H]$<$--1.6 dex) IACs with well populated RGBs prevents 
us from providing more firm constraints on Carina's intermediate-age population.    
These findings do not depend on the adopted metallicity scale, and if we 
adopt instead iron abundances on the \citet{car09} metallicity scale (see column 3 in Table~1), the 
metallicity range inferred for Carina is very similar. 

For a new estimate of the metal abundance of Carina 
we decided to use a new indicator. Evolutionary prescriptions for old 
and intermediate-age stellar structures indicate that the difference in color between 
the center of the RR Lyrae instability strip and the RC stars is strongly correlated 
with iron abundance. The correlation is caused by the mild dependence of the color 
of the instability strip on iron abundance and by the significant dependence 
of the color of RC stars when moving from metal-poor to metal-rich compositions 
(see Fig.~4). We estimated the predicted difference in color between  
RC stars (at the beginning of central helium burning) and the center of 
RR Lyrae strip ($\log T_e$=3.85) 
from the large set of scaled-solar evolutionary models \citep{pie04}  
available in the BaSTI database\footnote{Evolutionary models can be download 
from http://www.oa-teramo.inaf.it/BASTI}. 
We adopted evolutionary prescriptions that cover a wide range of iron abundances 
(--2.27$\le$[Fe/H]$\le$+0.06 dex), and an age range of 1--6 Gyr for the intermediate-age 
population and 13 Gyr for the old population. Note that the instability strip is 
minimally affected by cluster age when moving from 7--8 to 12--13 Gyr. In this context 
two robust evolutionary predictions need to be underlined. ({\em i}) 
When moving from metal-poor to metal-rich compositions, the intermediate-age
RC becomes  redder than 
the old RGB for [Fe/H]=--1.79 dex (see the top panel of Fig.~4). 
Current evolutionary models predict that a change in metallicity from 
[Fe/H]=--1.96 to [Fe/H]=--1.49 produces a change in $V$ magnitude of RC stars of 
0.03 mag, but a change of 0.12 mag in the \bmv\ color.  
({\em ii}) The spread in magnitude of HB stars at the mean color of the RR Lyrae 
strip increases, as expected, with the spread in metallicity. Note that a change in 
metallicity from [Fe/H]=--1.96 to [Fe/H]=--1.49 implies a change in $V$ magnitude of 
HB stars of 0.10 mag, but a change smaller than 0.01 mag in the \bmv\ color 
of the instability strip (see the bottom panel of Fig.~4).

Finally, we performed a linear regression between metallicity and the RC--HB difference
in three different colors:

\begin{tabular}{r l}
$[Fe/H]$ =&$ (-3.02\pm 0.03)+(3.89\pm 0.05)\Delta(\bmv)_{HB}^{RC}$ \\ 
$[Fe/H]$ =&$ (-3.47\pm 0.04)+(2.37\pm 0.04)\Delta(\bmi)_{HB}^{RC}$ \\ 
$[Fe/H]$ =&$ (-4.13\pm 0.08)+(6.02\pm 0.15)\Delta(\vmi)_{HB}^{RC}$ 
\end{tabular}

The above relations (see also right panels in Fig.~5) indicate that the difference in 
the \bmi\ color is significantly more sensitive to [Fe/H] than the \bmv\ color. The  
\vmi\ is the least sensitive of all and shows a stronger nonlinear trend in the 
metal-poor ([Fe/H]$\le$--1.5) regime.   
We estimated the difference in color between the RC and the middle of the 
RR Lyrae instability strip (see the left panels in Fig.~5). 
We found $\Delta$\bmv$^{RC}_{HB}$=0.37$\pm$0.05,  
$\Delta$\bmi$^{RC}_{HB}$=0.75$\pm$0.05 and $\Delta$\vmi$^{RC}_{HB}$=0.40$\pm$0.05 mag, 
where the errors account for uncertainties in the photometry and in the estimate of 
the mean color of the RR Lyrae strip. We applied the above relations and we found 
the following mean metallicities: 
[Fe/H]=--1.60$\pm$0.24($\Delta$\bmv$^{RC}_{HB}$),    
[Fe/H]=--1.70$\pm$0.19($\Delta$\bmi$^{RC}_{HB}$) and  
[Fe/H]=--1.76$\pm$0.44($\Delta$\vmi$^{RC}_{HB}$).
We adopted the estimate based on \bmi, since it is the most precise.

As a further test to constrain the spread in metallicity of the Carina 
sub-populations, we computed a series of synthetic CMDs changing the 
spread both in age and in metal content. The set of evolutionary models 
used for these numerical simulations are the same used to derive the 
new metallicity indicator (Pietrinferni et al.\ 2004). The code adopted 
to compute the synthetic CMDs has already been discussed by 
Pietrinferni et al.\ (2004) and by Cordier et al.\ (2007). We only briefly 
mention the initial parameters adopted to compute the synthetic CMDs. 
The Initial Mass Function was modeled using a power law with a Salpeter exponent 
($\alpha$=-2.35), and we assumed a fraction of unresolved binaries of the order of 
10\% with a minimum mass ratio for the binary systems equal to 0.7. 
The synthetic CMDs were constructed assuming two star formation events, 
with a mean age of 2.5--4 Gyr for the intermediate-mass and of 11--13 Gyr for 
the low-mass sub-population (Monelli et al.\ 2003). We also assumed that the two star 
formation episodes include the same fraction of stars. Moreover, we assumed mean 
metallicities of [Fe/H]=--1.5 and of --2.0 dex for the intermediate-age and the old 
burst, respectively. In accordance with the above findings, we assumed the same spread in 
metallicity---$\pm$0.1 dex---for the intermediate and the old star formation episode. 
Finally, to mimic actual observations, synthetic photometric errors were added using a 
Gaussian in the \B\ and \V\ bands with a dispersion equal to 0.02 mag. The above 
assumptions adopted to simulate the Carina observations are crude, but a 
detailed analysis is beyond the aims of this investigation. We plan to study the Carina 
star formation history in a forthcoming paper (Monelli et al.\ 2010). Data plotted in 
the top panel of Fig.~6 show that the aforementioned assumptions provide evolutionary features 
that are in reasonable agreement with the observations (see Fig.~2). In particular, 
both RC and HB stars are clearly separated in magnitude and in color from each other 
and from RG stars. Moreover, we selected two bins along the RGB for 
20.75$\le$$m_v$$\le$21.25 and 21.50$\le$$m_v$$\le$22.00 and we found that the spread 
in \bmv\ color is $\sim$0.06 mag. This value is quite similar, within the errors, to 
the spread in \bmv\ color of the observed CMD at the same magnitude intervals,
i.e., 0.06 mag. To constrain the impact of a spread in metal content, we used
the same evolutionary ingredients adopted to construct the above synthetic CMD,
but we assumed for each of the two stellar components a spread in metallicity of
0.2 dex.  The spread in \bmv\ color at the same magnitude intervals is slightly
larger than observed, namely 0.08 mag.    

Finally, we adopted for the intermediate-age stellar component alone a spread in metallicity 
of 0.5 dex. Data plotted in the bottom panel of Fig.~6 clearly show that such a spread 
in metallicity causes the RC stars to attain colors that are redder than the RG stars. 
The spread in \bmv\ color at the same magnitude intervals is a factor of two larger than
observed, i.e., 0.12 mag.
Moreover, the synthetic CMD suggests the occurrence of subgiant and giant branch stars 
with colors that are redder than the subgiant of the old sub-population. The observed 
CMD does not show these features.      

The comparison with empirical calibrators further supports the contention that 
Carina hosts predominantly metal-poor stellar populations, in agreement with 
the spectroscopic investigations. However, the limited range in color covered 
by the RC stars appears incompatible with the spectroscopic results: the spread 
in iron abundance of the intermediate-age population is either small, or is
counterbalanced by some other---as yet unrecognized---variable. The same
conclusion applies for the old population, based on the limited magnitude 
spread among the old HB stars. These conclusions are also supported by 
synthetic CMDs constructed assuming a different spread in metallicity between 
the intermediate and the old sub-populations. The simulations indicate a spread 
in metallicity of the two sub-populations smaller than 0.2 dex.


\acknowledgments
It is a real pleasure to thank an anonymous referee for his/her positive
comments on the results of this investigation and for his/her suggestion.
We also thank A. Grazian, for several useful discussions concerning 
the intrinsic color of galaxies as a function of the redshift and K. Glatt for 
sending us her photometric catalogs of the three SMC clusters. This project was 
supported by Monte dei Paschi di Siena (P.I.: S. Degl'Innocenti), PRIN-INAF2006 
(P.I.: M. Bellazzini).
HAS thanks the National Science Foundation for support under grant AST0607249.
 


\begin{deluxetable}{lccccc}
\tablewidth{0pt}
\tablecaption{Intrinsic parameters of old and intermediate-age calibrating clusters.}\label{tbl-1}
\tablehead{
\colhead{Name}&
\colhead{[Fe/H]}&
\colhead{[Fe/H]\tablenotemark{c}}&
\colhead{$\mu$}&
\colhead{E(\bmv)}&
\colhead{Age}
}
\startdata
NGC~1904 - M79 & -1.64$\pm$0.10\tablenotemark{a} & -1.58$\pm$0.02 & 15.61$\pm$0.10\tablenotemark{d} & 0.01$\pm$0.02\tablenotemark{d} & 12.6$\pm$1.3\tablenotemark{f}\\
NGC~6809 - M55 & -1.85$\pm$0.10\tablenotemark{a} & -1.93$\pm$0.02 & 13.69$\pm$0.10\tablenotemark{d} & 0.09$\pm$0.02\tablenotemark{d} & 12.4$\pm$1.7\tablenotemark{f}\\
NGC~5024 - M53 & -2.02$\pm$0.15\tablenotemark{a} & -2.06$\pm$0.09 & 16.24$\pm$0.10\tablenotemark{d} & 0.01$\pm$0.02\tablenotemark{d} & 11.2$\pm$1.8\tablenotemark{f}\\
Kron~3         & -1.08$\pm$0.12\tablenotemark{b} & -0.97$\pm$0.14 & 18.83\tablenotemark{e} & 0.006\tablenotemark{e} & 6.3\tablenotemark{e}\\
NGC~339        & -1.36$\pm$0.16\tablenotemark{b} & -1.25$\pm$0.18 & 18.75\tablenotemark{e} & 0.04\tablenotemark{e} & 6.0\tablenotemark{e}\\
Lindsay~38     & -1.59$\pm$0.10\tablenotemark{b} & -1.52$\pm$0.12 & 19.00\tablenotemark{e} & 0.04\tablenotemark{e} & 6.5\tablenotemark{e}\\
\enddata
\tablenotetext{a}{Iron abundance by \citet{kraft04}.}
\tablenotetext{b}{Iron abundance in the ZW84 metallicity scale by 
\citet[Kron~3, NGC~339]{dac98} and \citet[Lindasy~38]{kay07}.}
\tablenotetext{c}{Iron abundance by C09.} 
\tablenotetext{d}{Distance modulus and reddening by  
\citet[M79]{ferr99}, \citet[M55]{varg07} and \citet[M53]{dek09}. 
The uncertainty on distance and reddening are 5\% and 20\%.}
\tablenotetext{e}{Distance modulus, reddening and age by \citet{glatt09} 
using cluster isochrones by \citet{dot07}.}
\tablenotetext{f}{Cluster ages: \citet[M79, M55]{salaris02}, \citet[M53]{sar09}.}
\end{deluxetable}

\clearpage
\begin{figure}[!ht]
\begin{center}
\label{fig1}
\includegraphics[height=0.60\textheight,width=0.80\textwidth]{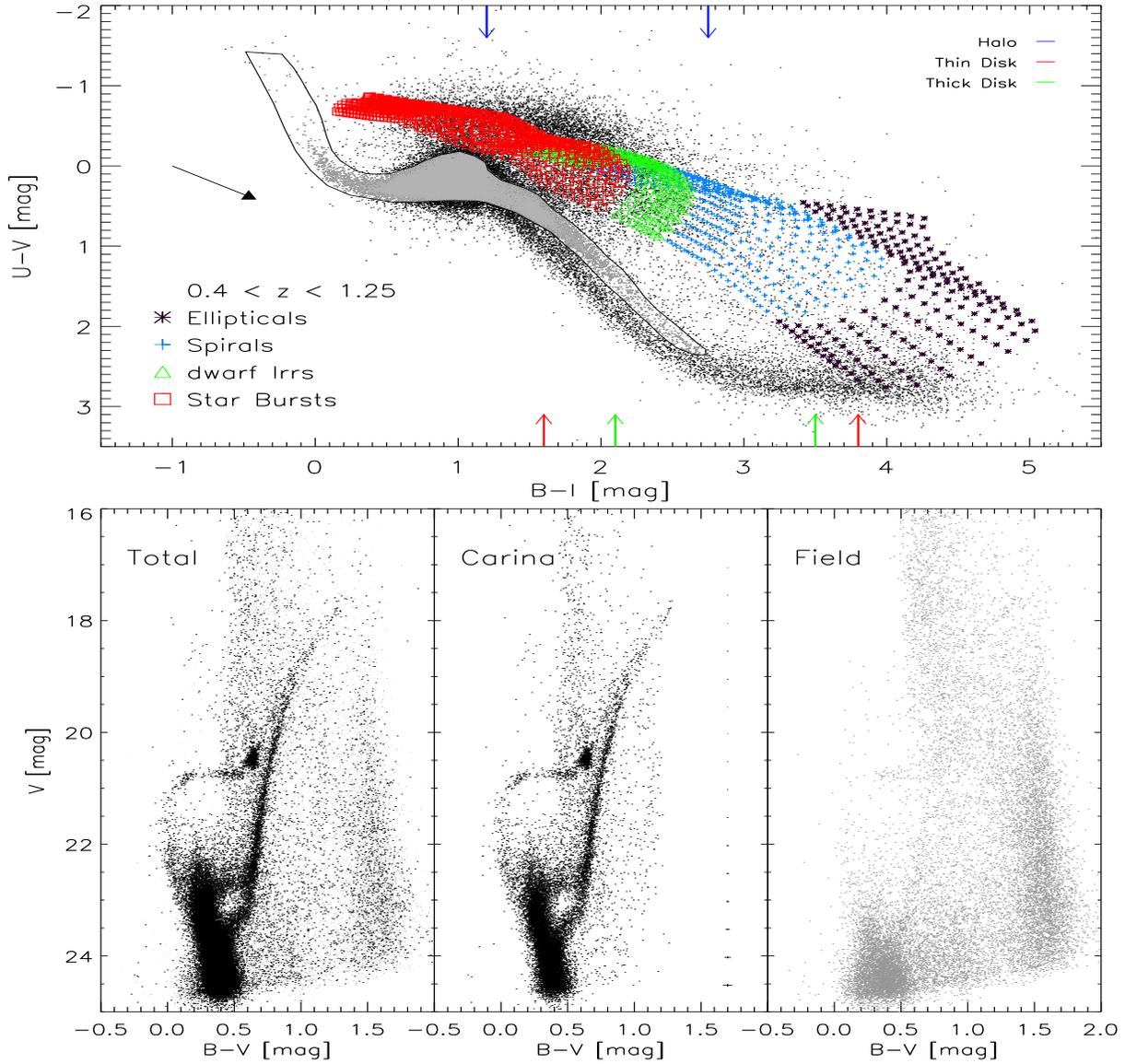}
\vspace*{0.75truecm}
\caption{Top -- $\umv,\bmi$ Color-Color Diagram (C-CD) of Carina dSph.  The curved
box encloses candidate Carina stars (grey dots). The black dots represent
probable field objects. Predicted distributions
\citep{rocca08} of elliptical (asterisks), spiral (pluses), SB (triangles), 
irregular and star burst (squares) galaxies as a function of the redshift are also showed. 
The vertical colored arrows mark the predicted \citep{cast02} peaks in \bmv\ color 
for the halo (blue), the thick disk (green) and the thin disk (red). The black arrow shows 
the reddening vector.
Bottom -- Left -- $V, \bmv$ Color-Magnitude Diagram (CMD) of Carina dSph.  Bottom --
Middle -- $V, \bmv$ CMD of candidate field stars according to the C-CD selection.  
The error bars on the right represent the intrinsic
photometric error in magnitude and in color.  Rejection of thick-disk turnoff stars
is obviously imperfect near \bmi$\sim$0.5, where the C-CD loses discriminating power.
Bottom -- Right -- $V$,\bmv\ CMD of
candidate Carina stars according to the C-CD selection.  A hint of the Carina HB 
near $V$$\approx$20.75 and of the SGB at 22.5$\ltsim$$V$$\ltsim$23.75 confirm that 
discrimination is imperfect at these intermediate colors.  The block of dots with 
\bmi$<$0.5, $V$$>$23.5 is probably dominated by background galaxies rather than 
Carina turnoff stars.
}        
\end{center}
\end{figure}

\begin{figure}[!ht]
\begin{center}
\label{fig2}
\includegraphics[height=0.70\textheight,width=0.95\textwidth, angle=90]{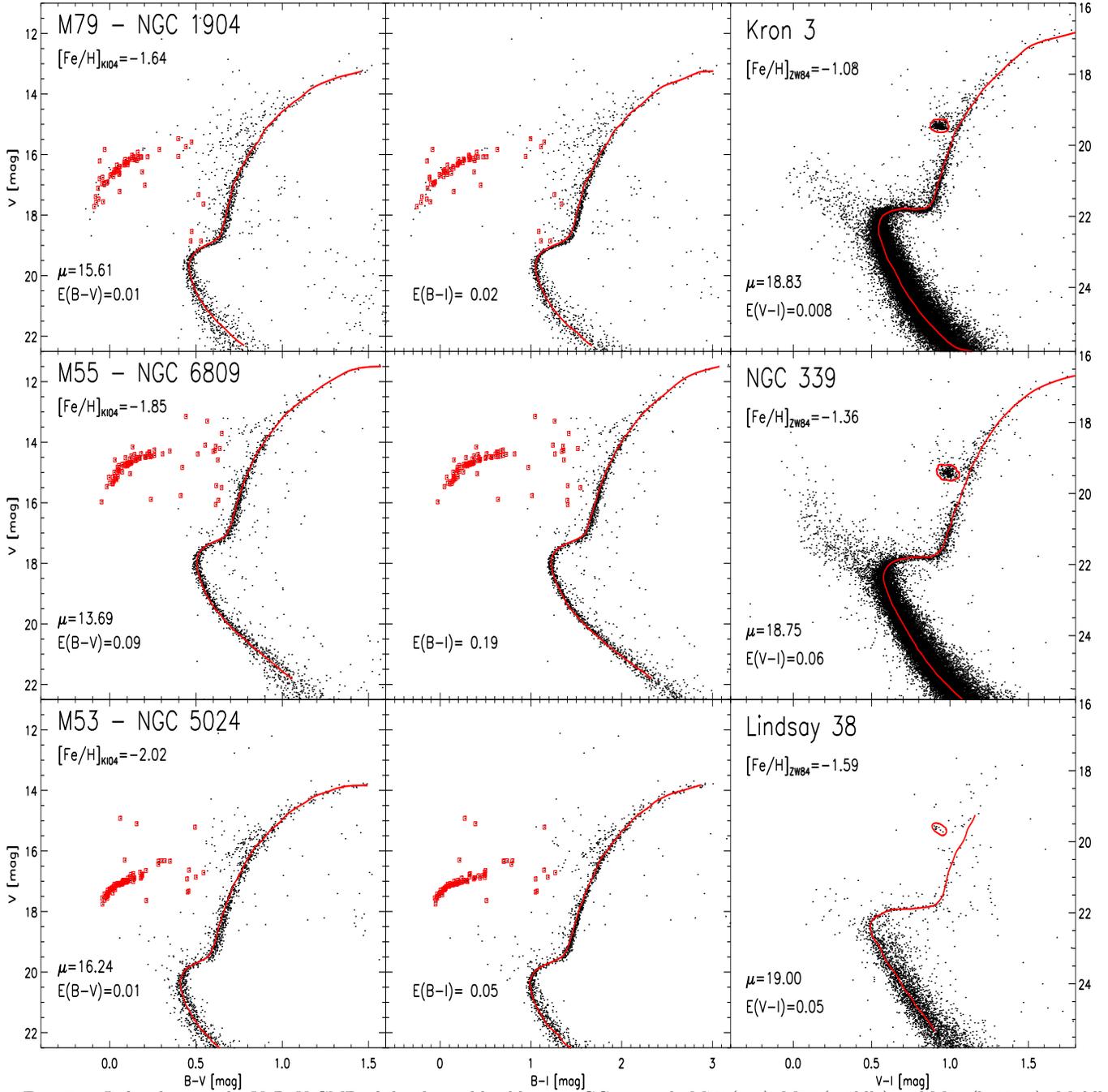}
\vspace*{0.5truecm}
\caption{Left column -- the $V, \bmv$ CMD of the three old calibrating GCs, namely 
M79 (top), M55 (middle) and M53 (bottom). Middle column -- same as the left, but for the 
$V, \bmi$ CMD. Right column -- the $V, \vmi$ CMD of the three intermediate-age SMC calibrating 
clusters, namely Kron~3 (top), NGC~339 (middle) and Lindsay~38 (bottom). The red lines 
display the ridge lines, the red squares mark old HB stars and the red oval the contour 
of RC stars. The iron abundance, the distance modulus, and the reddening are also labeled.    
}
\end{center}
\end{figure}

\begin{figure}[!ht]
\begin{center}
\label{fig3}
\includegraphics[height=0.70\textheight,width=0.95\textwidth, angle=90]{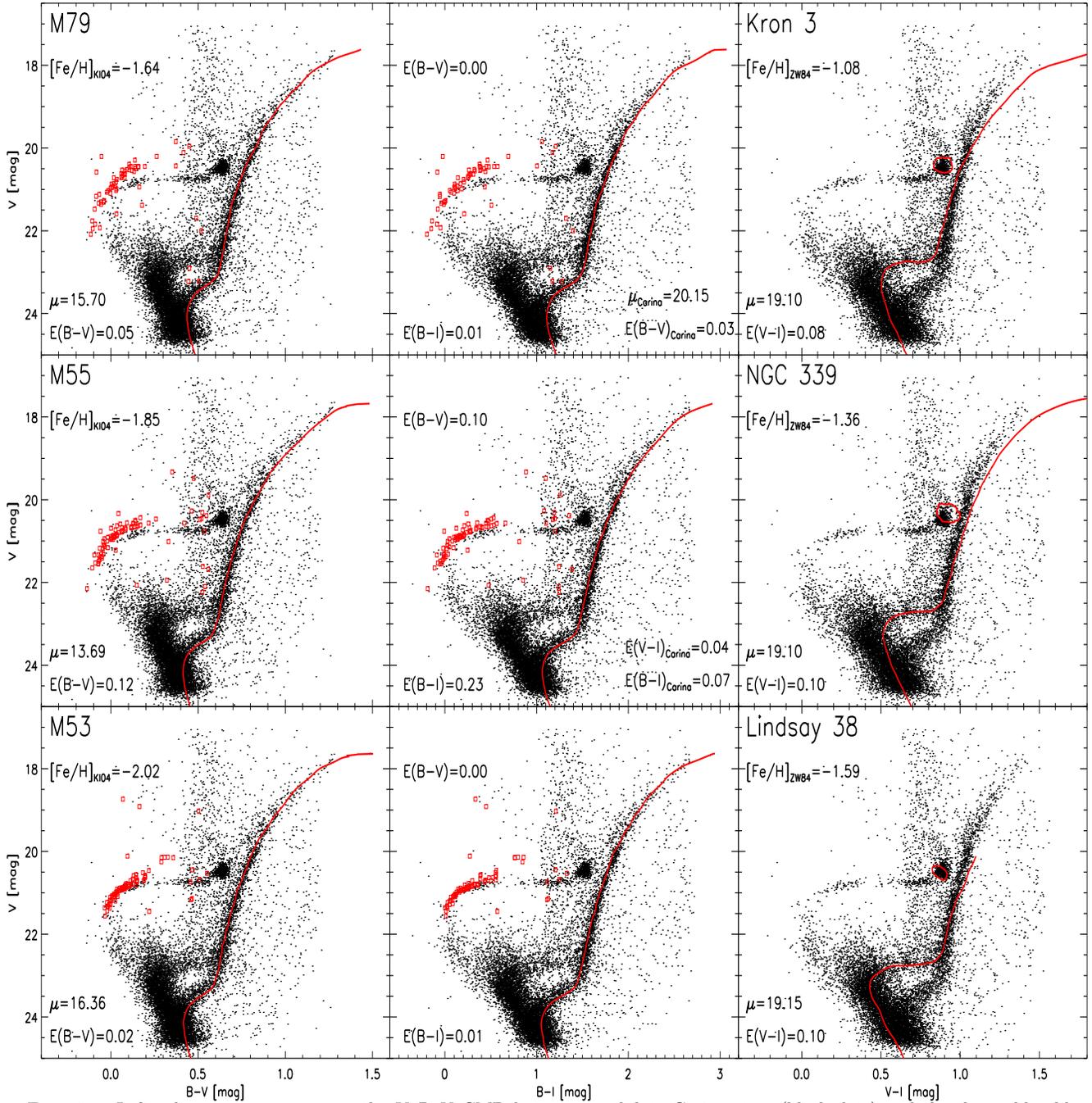}
\vspace*{0.5truecm}
\caption{Left column -- comparison in the $V, \bmv$ CMD between candidate Carina stars (black dots) and 
the three old calibrating GCs, namely M79 (top), M55 (middle) and M53 (bottom). The distance modulus 
and the reddening adopted to overlap the SGB of old GCs with the SGB of Carina are labelled. 
Middle column -- same as the left, but in the $V, \bmi$ CMD. Right column -- comparison in the $V, \vmi$ CMD 
between Carina stars and the three intermediate-age SMC calibrating clusters, namely Kron~3 (top), 
NGC~339 (middle) and Lindsay~38 (bottom). The distance modulus and the reddening adopted to overlap 
the SGB of intermediate age SMC clusters with the SGB of Carina are labelled. 
}
\end{center}
\end{figure}

\begin{figure}[!ht]
\begin{center}
\label{fig4}
\includegraphics[height=0.60\textheight,width=0.55\textwidth]{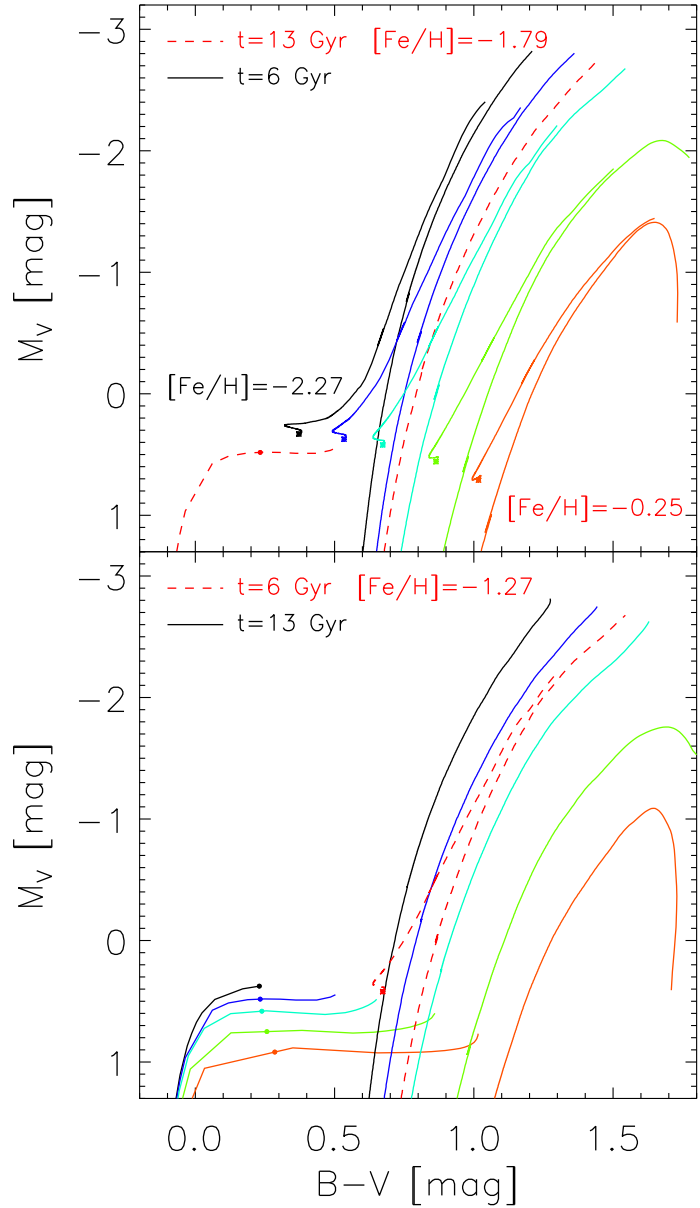}
\vspace*{1.55truecm}
\caption{
Top -- cluster isochrones in the $V, \bmv$ CMD at fixed age (6 Gyr) 
and for a broad range of chemical compositions (BaSTI database). The vertical red 
dashed line shows the isochrone for an old (13 Gyr), metal-poor ([Fe/H]=--1.79 dex) stellar 
structure together with its ZAHB. The colored lozenges mark the RC color, while the red square 
marks the middle of the RR Lyrae instability strip ($\log T_e$=3.85). 
Bottom -- cluster isochrones in the $V, \bmv$ CMD at fixed age (13 Gyr) and for a broad range of 
chemical compositions (BaSTI database). The vertical red dashed line shows the isochrone for 
an intermediate age (6 Gyr), metal-intermediate ([Fe/H]=--1.27 dex) stellar structure together 
with its He-burning phases. The symbols are the same as in the top panel.      
}
\end{center}
\end{figure}

\begin{figure}[!ht]
\begin{center}
\label{fig5}
\includegraphics[height=0.65\textheight,width=0.50\textwidth]{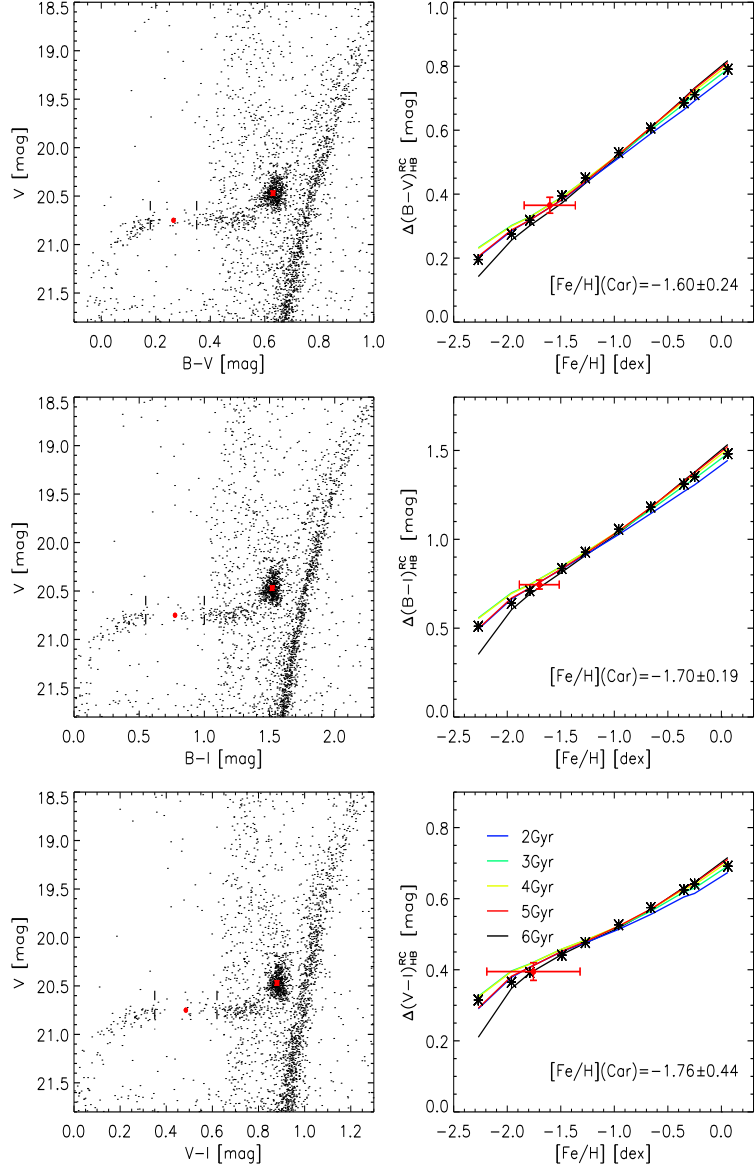}
\vspace*{0.5truecm}
\caption{
Left column -- zoom across the RC and old HB region of Carina in the $V, \bmv$ (top),  
$V, \bmi$ (middle) and in the $V, \bmi$ (bottom) CMD. The red circle marks the position 
of the middle of the RR Lyrae instability strip and the red asterisk marks the position 
of the peak of RC stars. 
Right column -- Predicted difference in \bmv\ (top), in \bmi\ (middle) and in \vmi\ (bottom) 
colors between the middle of the RR Lyrae instability strip and the peak of RC stars as a 
function of iron abundance.  
Lines of different colors display predictions for different ages of intermediate-mass stars 
(see labeled values). The black asterisks mark the BaSTI grid points. The red dots display 
Carina, while the error bars the uncertainties affecting the difference in color and the 
metallicity estimate.   
}
\end{center}
\end{figure}

\begin{figure}[!ht]
\begin{center}
\label{fig6}
\includegraphics[height=0.65\textheight,width=0.50\textwidth]{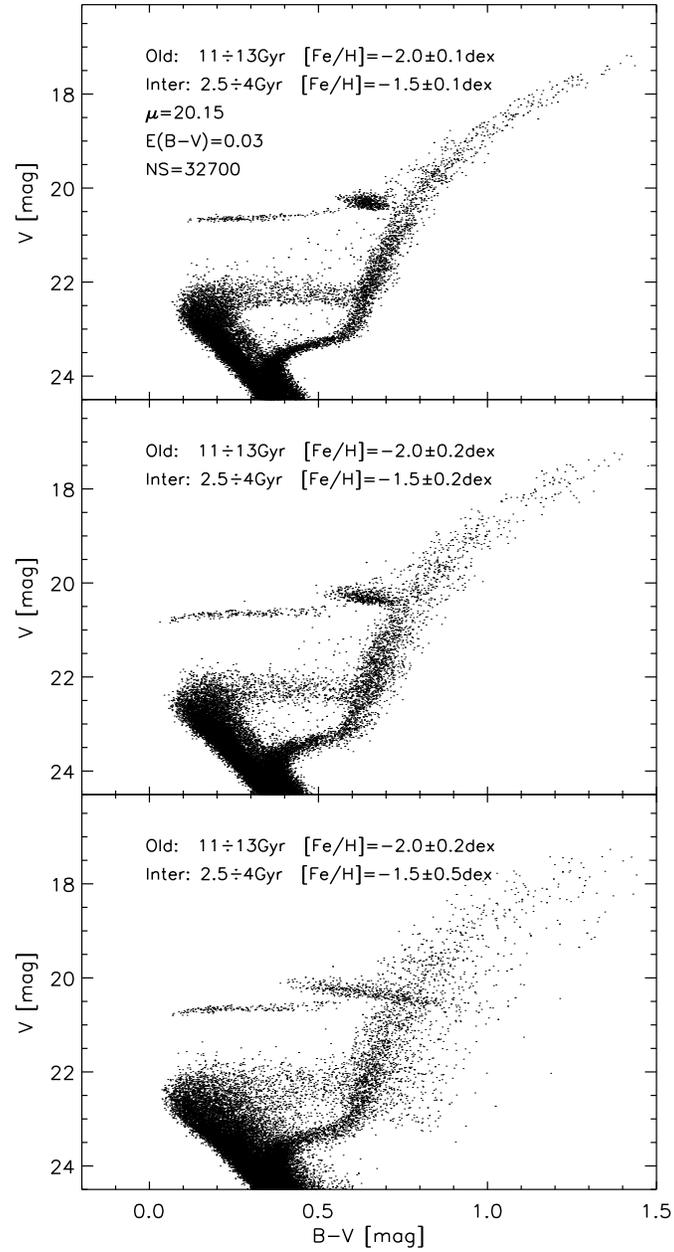}
\vspace*{1.5truecm}
\caption{ 
Top -- synthetic $V, \bmv$ CMD of Carina. The CMD was constructed by assuming 
two star formation episodes, with ages of 11$\div$13 Gyr for the old and of 
2.5$\div$4 Gyr for the intermediate-age stellar component (Monelli et al.\ 2003). 
The two star formation episodes include the same fraction of stars and  
have a mean metallicity of [Fe/H]=--1.5 and of --2.0 dex, respectively. 
The spread in metallicity is $\pm$0.1 dex for both the intermdiate and the 
old sub-population. Intrinsic photometric error was added using a Gaussian 
with a mean equal to 0.02 mag. 
Middle -- same as the top, but the synthetic CMD was constructed assuming a 
spread in metallicity of $\pm$0.2 dex for the old and the intermdiate-age 
sub-population.  
Bottom -- same as the middle, but the synthetic CMD was constructed assuming a 
spread in metallicity of $\pm$0.5 dex for the intermdiate-age sub-population.    
}
\end{center}
\end{figure}

\end{document}